# Understanding heavy fermion from generalized statistics

Y. Kaupp[a], S. Liraki[a], Dmitrii Tayurskii[b], Arthur Useinov[b], A. El Kaabouchi[a], L. Nivanen[a], B. Minisini[a], F. Tsobnang[a], A. Le Méhauté[a], Q. A. Wang[a]

[a]Institut Supérieur des Matériaux et Mécaniques Avancés du Mans, 44, Avenue F.A. Bartholdi, 72000 Le Mans, France

[b]Departement of physics, Kazan State University, Kazan 420008, Russia

## Abstract

Heavy electrons in superconducting materials are widely studied with the Kondo lattice *t-J* model. Numerical results have shown that the Fermi surface of these correlated particles undergoes a flattening effect according to the coupling degree *J*. This behaviour is not easy to understand from the theoretical point of view within standard Fermi-Dirac statistics and non-standard theories such as fractional exclusion statistics for anyons and Tsallis nonextensive statistics. The present work is an attempt to account for the heavy electron distribution within incomplete statistics (IS) which is developed for complex systems with interactions which make the statistics incomplete such that $\sum_{i=1}^{w} p_i^q = 1$. The parameter *q*, when different from unity, characterizes the incompleteness of the statistics. It is shown that the correlated electrons can be described with the help of IS with *q* related to the coupling constant *J* in the context of Kondo model.

Over the last decades, statistical mechanics has experienced a rapid development of several extended theories for the equilibrium or nonequilibrium systems having interacting elements or complex behaviors such as formation of fractal structure by phase space trajectories. These theories are intended to describe non standard properties such as nonextensivity[1][2][3], mixture of boson-fermion properties of quasi particle with fractional Pauli exclusion[7][8][9], quantum behaviors with extended commutation rules as in quantum group theory[5][6], and systems having complex interaction and correlation leading to incompleteness of statistical description[10][11][12]. A common character of these extensions is the introduction of empirical parameters in order to take into account the effects out of the realm of the conventional statistical mechanics which is nevertheless recovered when the parameters take special values. For some of these theories, the empirical parameters have clear and precise physical meanings. For instance, in the fractional exclusion statistics (FES) [7][8][9] whose occupation number $n$ of fermion is given by the following distribution:

$$n = \frac{1}{f\left(e^{-\beta(e-e_f)}\right) + \alpha} \quad (0 \leq \alpha \leq 1) \tag{1}$$

where $f(x)$ is a function satisfying $f^\alpha(x)[1+f(x)]^{1-\alpha} = x$, or equivalently by[13]

$$n = \frac{1}{e^{\beta(e-e_f)} - 1} - \frac{\frac{1+\alpha}{\alpha}}{e^{\frac{1+\alpha}{\alpha}\beta(e-e_f)} - 1} \quad (0 \leq \alpha \leq \infty) \tag{2}$$

Note that in this version the parameter $\alpha$ varies between zero and infinity and can be related to the maximal occupation number of a quantum state by $n_{\max} = 1/\alpha$. In the nonextensive incomplete statistics (NIS) with a fermion distribution such as[15]

$$n = \frac{1}{\left[1 + (q-1)\beta(e-e_f)\right]^{\frac{q}{q-1}} + 1}, \tag{3}$$

where the parameter $q$ ($0 < q < \infty$) can be related to the missing physical states not taken into account in the normalization[^1] or, for nonequilibrium system evolving in phase space in multifractal attractors, to the similarity dimension of the attractors[14]. In the extensive version of incomplete statistics (EIS)[12], the fermion distribution is given by

[^1]: The incomplete normalization is characterized by $\sum_{i=1}^{w} p_i^q = 1$. When $q=1$, the normalization recovers the conventional one.

$$n = 1/(e^{-q\beta(e-e_f)} + 1). \tag{4}$$

And finally in the nonextensive statistical mechanics (NSM) of Tsallis[3], the fermion distribution is given by[16]

$$n = \frac{1}{[1 + (q-1)\beta(e - e_f)]^{\frac{1}{q-1}} + 1}. \tag{5}$$

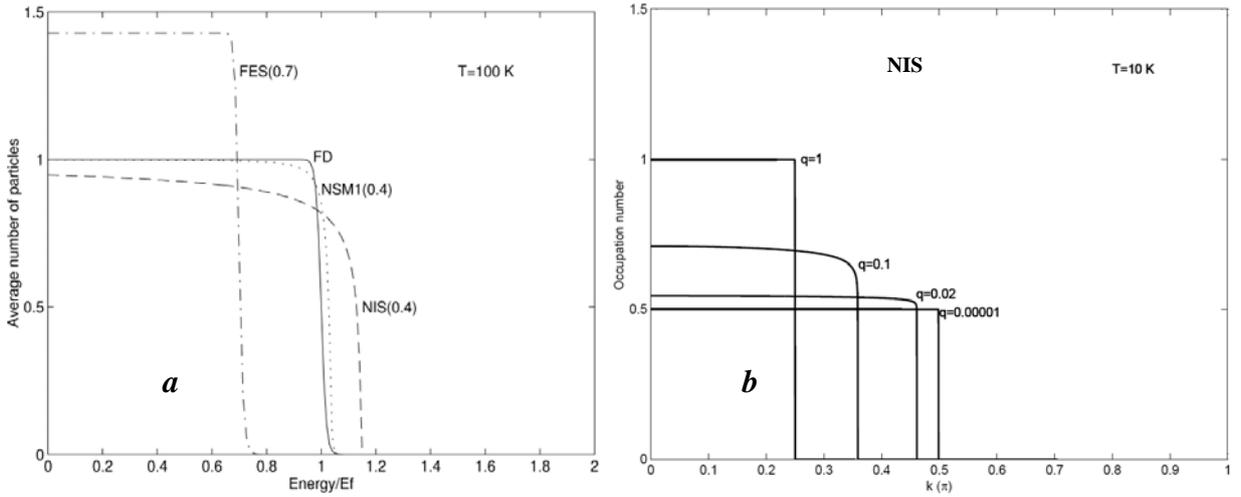

**Figure 1**: (**1a**) Fermion distributions of Fermi-Dirac (FD), FES ($\alpha$ = 0.85), NSM ($q$ = 0.4) and of nonextensive IS (NIS, also with $q$ = 0.4) at 100 $K$. The fermion density is chosen to give $e_f^0$ = 1 $eV$ for FD distribution at $T$ = 0. We note that all these distributions show sharp $n$ drop at Fermi energy $e_f$. Within FES, this sharp drop does not change with $\alpha$. Similar for NSM distribution which is only slightly different from FD one even with $q$ very different from unity. (**1b**) On the contrary, NIS distribution changes drastically with decreasing $q$, showing a wide decrease of $n$ and a strong increase of $e_f$ with decreasing $q$. The fermion density is chosen to give Fermi momentum $k_f^0 = 0.25\pi$ at $T$ = 0 and $q$=1.

To give the reader an idea about the behaviors of the distribution mentioned above, the fermion distributions of Fermi-Dirac (FD) statistics, FES, NSM and NIS are shown in Figure 1 for low temperatures $T$=100 $K$ and $T$=10 $K$. We note the sharp drop in occupation number $n$ around Fermi energy $e_f$ at low temperature. The drop rate suffers very little the influence of the variation in $q$ and $\alpha$. This variation leads mainly to the change in the Fermi energy $e_f$. We also note that NSM fermion distribution has a very little change of Fermi energy with $q$. So the entire distribution remains almost the same for whatever $q$ value. However, the NIS Fermi energy increases tremendously with decreasing $q$ with a wide decrease in $n$ between zero energy to the Fermi energy $e_f$ (or Fermi momentum $k_f = \sqrt{2me_f}/\hbar$, see Figure 1b). A common character of NSM and NIS is that the electrons are almost forbidden to overcome the Fermi energy due to the energy cutoff, $1+(q-1)\beta(e-e_f) > 0$ in Eqs.(3) and (5).

This present work is the continuation of our previous effort to account for the behaviour of correlated heavy electrons observed in numerical simulations with the one-dimensional Kondo lattice model (KLM)[23][24][25]. These electrons have many singular properties which cannot be accounted for with conventional FD theory. One of these anomalous properties is the flattening of the steep drop of occupation number around the Fermi energy with increasing correlation at low temperatures, as shown in Figure 2a (symbols).

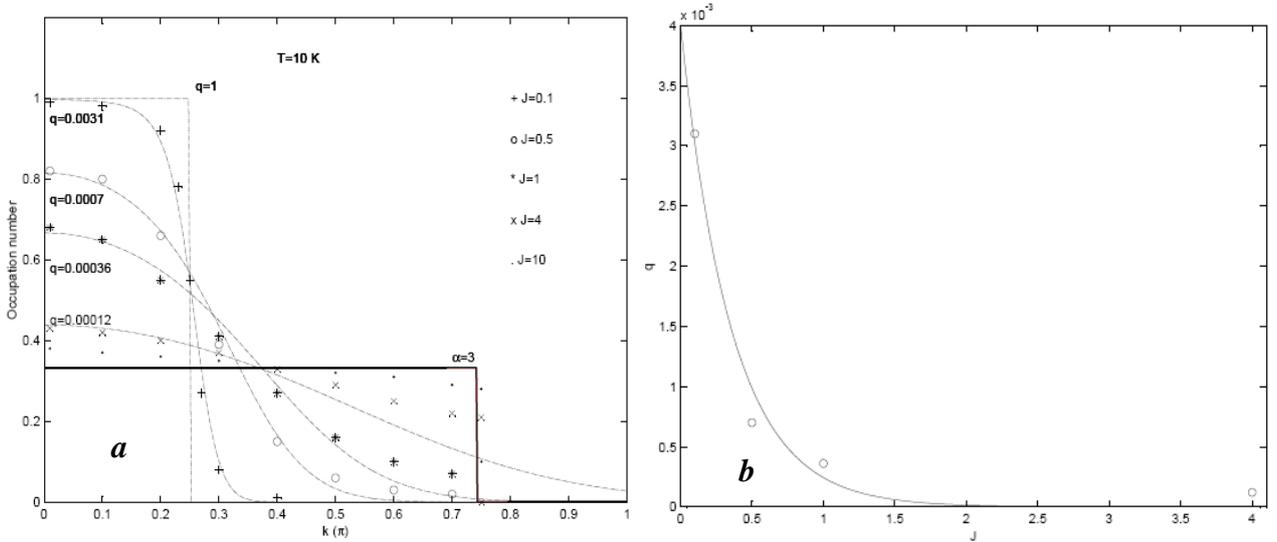

**Figure 2**: (**2a**) : Comparison of EIS fermion distribution (dotted lines) with the numerical results (symbols) of Eder el al on the basis of Kondo lattice $t-J$ model (KLM)[23][24]. The density of electrons is chosen to give $k_f^0 = 0.25\pi$ in the first Brillouin zone. We note that EIS distribution reproduces well the numerical results for about $J < 1$. When the coupling is stronger, a fat tail in the KLM distributions begins to develop at higher energy and a new Fermi surface at $k = k_f^0 + \pi/2 = 0.75\pi$ starts to appear with a sharp $n$ drop. EIS fails to account for this property since it has a long tail without cutoff in energy or momentum $k$. (**2b**) : Relation between $J$ and $q$ up to $J=4$ (symbols) which can be fitted by $q = c\,e^{-\gamma J}$ (full line) with $c = 4\times 10^{-3}$ and $\gamma = 2.8$.

In the KLM model, the correlation is characterized by a parameter $J$. When there is no correlation, the Fermi momentum (vector) is $k_f^0 = 0.25\pi$ in the calculations of [23] and [24]. When the correlation is relatively weak with $J$ smaller than unity, the $n$ drop around $e_f$ begins to flatten with a small increase of Fermi vector $k_f$. This property can be accounted for with EIS distribution Eq. (4) as shown by the dotted lines in Figure 2a which fit satisfactorily the results of numerical simulation of KLM. We see that the increasing $J$ corresponds to decreasing $q$. The relation between J and q can be represented by an empirical formula such as $q = 4 \times 10^{-3} e^{-2.8J}$ as shown in Figure 2b. This result can be interpreted as follows. The ideal gas model becomes statistically incomplete due to the interactions introduced in the KLM model. This incompleteness can be to some extent represented by the parameter $q$ of Eq. (4).

However, for stronger correlation with $J$ larger than unity[23][24], a long tail in the KLM distributions begins to develop at higher energy. At the same time, a new Fermi momentum at $k_f = k_f^0 + \pi/2 = 0.75\pi$ starts to appear and a sharp $n$ drop takes place at the new Fermi momentum. Due to the new Fermi energy which is about three times $k_f^0$, the occupation number becomes lower than $n=1/2$ and tends to a uniform distribution between $k=0$ and the new Fermi vector $k_f$ (Figure 2). This behavior cannot be understood with Eq. (4) of EIS, since EIS can have an increasing Fermi vector with decreasing $q$, but it cannot have a sharp cutoff of occupation number at higher vector. NSM's Fermi energy is almost constant, which is not the case of Figure 2. Finally, NIS can have increasing Fermi vector. But the upper limit of $k_f$ is two times $k_f^0$ since the occupation number cannot be smaller than 0.5 corresponding to $q=0$ (see Figure 1b).

As for FES, it cannot have the flattening of occupation number for fixed temperature so it is not suitable for the case of weak interaction. On the contrary, it is possible to use it to reproduce the cutoff distribution of strong interaction (around $J=10$) which tends to yield a uniform distribution with $n$ close to roughly 0.35 and a Fermi vector around $0.75\pi$, a constant value independent of $J$. This kind of distribution implies that the maximal occupation number is smaller than one. This is a characteristic of the FES given by Eq.(2) with $0 < \alpha < \infty$. In Figure 2a, the full line represents the limiting distribution for strong correlation given by Eq.(2) with $\alpha = 3$.

On the other hand, it is interesting to note the property of the fermion distribution of FES within NIS which has not been useful in the present study of KLM correlation. We present it here just to show their unexpected property. This distribution can be derived through the conventional method[13][15] and is given by

$$n = \frac{1}{\left[1+(q-1)\beta(e-\mu)\right]^{\frac{q}{q-1}}-1} - \frac{\frac{1+\alpha}{\alpha}}{\left[1+(q-1)\frac{1+\alpha}{\alpha}\beta(e-\mu)\right]^{\frac{q}{q-1}}-1}. \tag{6}$$

In view of the behaviour of Eq.(3) of NIS, it may be expected that the decreasing $q$ would lead to decreasing occupation number and increasing Fermi vector. But the distribution shows the contrary (see figure 3). When $q$ decreases for fixed $\alpha$, the occupation number increases and the Fermi vector decreases with more and more fermions concentrated around the Fermi energy. We may speculate that this concentration might leads to the increase of conductivity with decreasing $q$. But for the moment, we have not seen any physical relevance of this property.

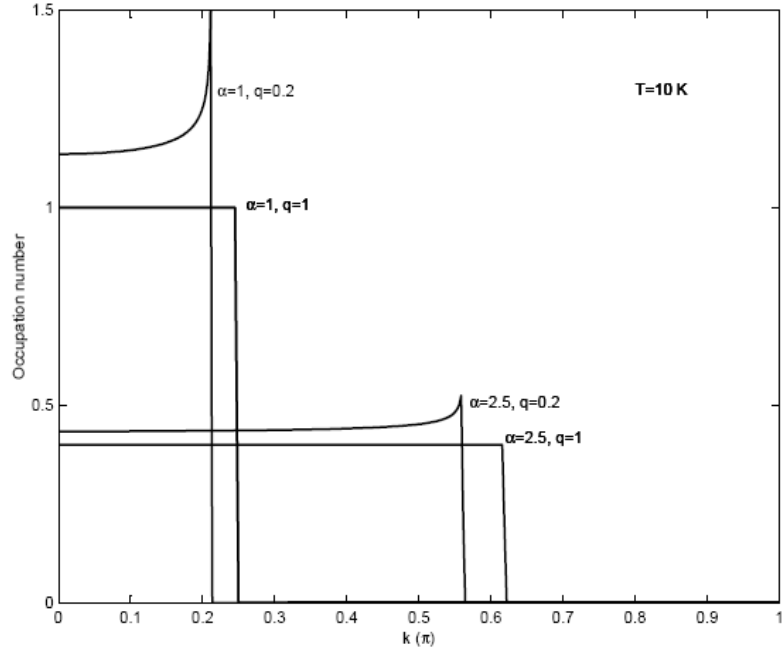

**Figure 3**: Fermion distribution of FES with NIS.

To conclude this work, the correlated electrons in the KLM model undergo two different regimes with completely different statistical properties: the weak coupling regime and the strong coupling one. The weak coupling regime ($J<1$) can be satisfactorily described within EIS whose incompleteness can be described by the parameter $q$ related to the coupling parameter $J$ by $q = c\,e^{-\gamma J}$ (full line) with $c = 4\times 10^{-3}$ and $\gamma = 2.8$. But EIS fails for the strong coupling regime due to the absence of $n$ cutoff at high energy. The strong coupling case ($J>10$) has a limit distribution which can be reproduced within FES, i.e., there is a fractional Pauli exclusion which makes it possible to understand the maximal occupation number different from unity as shown in Figure 2*a*. However, the transition behaviors (1<*J*<10) between the two regimes cannot be exactly reproduced by the statistics used in this work. As far as we know, there is actually no theory that can satisfactorily yield these transition regimes observed in KLM model.